\documentclass[aps,prd,twocolumn,groupedaddress,amssymb,eqsecnum,showpacs,epsfig]{revtex4}
\usepackage{graphicx}
\usepackage{bm}
\usepackage{dcolumn}
\usepackage{amsmath}
\def \lleq {\lower0.9ex\hbox{ $\buildrel < \over \sim$} ~}
\def \ggeq {\lower0.9ex\hbox{ $\buildrel > \over \sim$} ~}

\newcommand{\rd}{{\rm d}}

\def \orad   {\Omega_{0r}}
\def \ol   {\Omega_{\Lambda}}
\def \ode   {\Omega_{DE}}

\def \omms   {\Omega_m}

\def \omr   {\Omega_r}
\def \omm  {\Omega_{0 {\rm m}}}

\def \beq  {\begin{equation}}
\def \eeq  {\end{equation}}
\def \ber  {\begin{eqnarray}}
\def \eer  {\end{eqnarray}}

\begin{document}
\newcommand{\newc}{\newcommand}

\newc{\be}{\begin{equation}}
\newc{\ee}{\end{equation}}
\newc{\ba}{\begin{eqnarray}}
\newc{\ea}{\end{eqnarray}}
\newc{\bea}{\begin{eqnarray*}}
\newc{\eea}{\end{eqnarray*}}
\newc{\D}{\partial}
\newc{\ie}{{\it i.e.} }
\newc{\eg}{{\it e.g.} }
\newc{\etc}{{\it etc.} }
\newc{\etal}{{\it et al.}}
\newcommand{\nn}{\nonumber}
\newc{\ra}{\rightarrow}
\newc{\lra}{\leftrightarrow}
\newc{\lsim}{\buildrel{<}\over{\sim}}
\newc{\gsim}{\buildrel{>}\over{\sim}}
\title{Can $f(R)$ Modified Gravity Theories Mimic a $\Lambda$CDM Cosmology?}
\author{S. Fay$^a$, S. Nesseris$^b$ and L. Perivolaropoulos$^b$}
\affiliation{$^a$School of
Mathematical Sciences, Queen Mary, University of London, UK
\\$^b$Department of Physics, University of Ioannina, Greece}
\date{\today}

\begin{abstract}
We consider $f(R)$ modified gravity theories in the metric variation
formalism and attempt to reconstruct the function $f(R)$ by
demanding a background $\Lambda$CDM cosmology. In particular we
impose the following requirements: a. A background cosmic history
$H(z)$ provided by the usual flat $\Lambda$CDM parametrization
though the radiation ($w_{eff}=1/3$), matter ($w_{eff}=0$) and
deSitter ($w_{eff}=-1$) eras. b. Matter and radiation dominate
during the `matter' and `radiation' eras respectively i.e. $\omms =1
$ when $w_{eff}=0$ and $\omr=1$ when $w_{eff}=1/3$. We have found
that the cosmological dynamical system constrained to obey the
$\Lambda$CDM cosmic history has four critical points in each era
which correspondingly lead to four forms of $f(R)$. One of them is
the usual general relativistic form $f(R)=R-2\Lambda$. The other
three forms in each era, reproduce the $\Lambda$CDM cosmic history
but they do not satisfy requirement b. stated above.
\end{abstract}
%
%
\maketitle

\section{Introduction}
There is accumulating observational evidence based mainly on Type Ia
supernovae standard candles \cite{SN} and also on standard rulers
\cite{CMB,BAO} that the universe has entered a phase of accelerating
expansion at a recent cosmological timescale. This expansion implies
the existence of a repulsive factor on cosmological scales which
counterbalances the attractive gravitational properties of matter on
these scales. There have been several theoretical approaches
\cite{CST,review} towards the understanding of the origin of this
factor. The simplest such approach assumes the existence of a
positive cosmological constant which is small enough to have started
dominating the universe at recent times. The predicted cosmic
expansion history in this case (assuming flatness) is \be H(z)^2 =
\left(\frac{{\dot a}}{a}\right)^2 = H_0^2 \left[\omm (1+z)^3 + \orad
(1+z)^4 + \ol\right] \label{lcdmz} \ee where
$\orad=\frac{\rho_{rad}}{\rho_{crit}}\simeq 10^{-4}$ is the present
energy density of radiation normalized over the critical density for
flatness $\rho_{crit}$. Also
$\omm=\frac{\rho_{m}}{\rho_{crit}}\simeq 0.3$ is the normalized
present matter density and $\ol=1-\omm-\orad$ is the normalized
energy density due to the cosmological constant. This model provides
an excellent fit to the cosmological observational data \cite{CMB}
and has the additional bonus of simplicity and a single free
parameter. Despite its simplicity and good fit to the data this
model fails to explain why the cosmological constant is so
unnaturally small as to come to dominate the universe at recent
cosmological times. This fine tuning problem is known as the {\it
coincidence problem}.

In an effort to address this problem two classes of models have been
proposed: The first class assumes that general relativity is a valid
theory on cosmological scales and attributes the accelerating
expansion to a {\it dark energy} component which has repulsive
gravitational properties due to its negative pressure. The role of
dark energy is usually played by a minimally coupled to gravity
scalar field called {\it quintessence\cite{quin}}. Alternatively,
the role of dark energy can be played by various perfect fluids (eg
Chaplygin gas \cite{chapgas}), topological defects \cite{defectsde},
holographic dark energy \cite{holographic} etc. The second class of
models attributes the accelerating expansion to a modification of
general relativity on cosmological scales which converts gravity to
a repulsive interaction at late times and on cosmological scales.
Examples of this class of models include scalar-tensor
theories\cite{BEPS00,stensor}, $f(R)$ modified gravity
theories\cite{fRpapers}, braneworld models \cite{braneworld} etc. An
advantage of models in this class is that they naturally
allow\cite{Perivolaropoulos:2005yv,BEPS00} for a superaccelerating
expansion of the universe where the effective dark energy equation
of state $w=\frac{p}{\rho}$ crosses the phantom divide line $w=-1$.
Such a crossing is consistent with current cosmological
data\cite{Alam:2003fg}.

Most of the models in both classes require the existence of
arbitrary new degrees of freedom whose role is usually played by
effective scalar fields. This is not a welcome feature because the
degrees of freedom are to some extend arbitrary with respect to
either their origin and/or their dynamical properties. Their
predictive power is therefore usually dramatically diminished.

A partial exception to this rule is provided by modified $f(R)$
theories of gravity. In these theories the Ricci scalar $R$ in the
general relativistic Lagrangian is replaced by an arbitrary function
$f(R)$ leading to an action of the form \be S=\int{\rm
d}^{4}x\sqrt{-g}\left[\frac{1}{2}f(R)+{\mathcal{L}}_{{\rm
rad}}+{\mathcal{L}}_{{\rm m}}\right]\label{actfr}\ee where
${\mathcal{L}}_{{\rm m}}$ and ${\mathcal{L}}_{{\rm rad}}$ are the
Lagrangian densities of matter and radiation and we have set $8\pi
G=1$.  These $f(R)$ theories arise in a wide range of different
frameworks: In quantum field theories in curved
spacetime\cite{birrell}, in the low energy limit of the $D=10$
superstring theory\cite{ON-Mth}, in the vacuum action for the Grand
Unified Theories (GUTs) etc.

It has been demonstrated\cite{Carroll} that for appropriate forms of
$f(R)$ the action (\ref{actfr}) can naturally produce accelerating
expansion at late times in accordance with SnIa
data\cite{noi-ijmpd}. The advantage of these theories is that no
extra arbitrary degree of freedom is introduced and the accelerating
expansion is produced by the Ricci scalar (dark gravity) whose
physical origin is well understood. On the other hand, the main
disadvantage of these theories is that (like most modified gravity
theories) they are seriously constrained by local gravity
experiments \cite{Chiba,Dolgov,Faraoni}. In fact it can be
shown\cite{Chiba} that $f(R)$ models are equivalent to scalar-tensor
theories with vanishing Brans Dicke parameter ($\omega=0$) and a
special type of potential. Since solar system tests of general
relativity imply $\omega>4\times 10^4$ \cite{pitjeva}, these
theories can only be consistent with observations if they are
associated with a large (infinite) effective mass of the scalar
 $R$. It has been shown \cite{NO03} that specific forms of the
function $f(R)$ can provide an infinite effective mass needed to
satisfy solar system constraints and can also produce late time
accelerating expansion.

The reduction of $f(R)$ theories to a special class of scalar-tensor
theories implies that in principle the reconstruction of $f(R)$ from
a particular cosmic history $H(z)$ can be performed in a similar way
as in the case of scalar-tensor
theories\cite{BEPS00,Perivolaropoulos:2005yv}. However, the
non-existence of a Brans Dicke parameter requires some modifications
of the reconstruction methods especially when the reconstruction
extends through the whole cosmic history through the radiation and
matter eras. The dynamical systems approach followed in the present
study illustrates these modifications

The construction of cosmologically viable models incorporating late
accelerating expansion based on $f(R)$ theories has been an issue of
interesting debate during the past year. This debate originated from
Ref. \cite{APT} which demonstrated that $f(R)$ theories that behave
as a power of $R$ at large or small $R$ are not cosmologically
viable because they have the wrong expansion rate during the matter
dominated era ($a\sim t^{1/2}$ instead of $a\sim t^{2/3}$). This
conclusion was challenged in Ref. \cite{Nojiri:2006gh} which claimed
that wide classes of $f(R)$ gravity models including matter and
acceleration phases can be phenomenologically reconstructed by means
of observational data. The debate continued with the recent Ref.
\cite{Amendola:2006we} where a detailed and general dynamical
analysis of the cosmological evolution of $f(R)$ theories was
performed. It was shown that even though most functional forms of
$f(R)$ are not cosmologically viable due to the absence of the
conventional matter era required by data, there are special forms of
$f(R)$ that can be viable (consistent with data) for appropriate
initial conditions.

In the present study we perform a generic model independent analysis
of $f(R)$ theories. Instead of specifying various forms of $f(R)$
and finding the corresponding cosmological dynamics, we specify the
cosmological dynamics to that of the $\Lambda$CDM cosmology and
search for a possible corresponding form of $f(R)$. We thus attempt
to {\it reconstruct} $f(R)$ from the background cosmological
dynamics. In particular we consider the general autonomous system
for cosmological dynamics of $f(R)$ theories and study the dynamics
of $f(R)$ using as input a $\Lambda$CDM cosmic expansion history.
Our study is performed both analytically (using the critical points
and their stability) and numerically by explicitly solving the
dynamical system. The results of the two approaches are in good
agreement since the numerical evolution of $f(R)$ follows the
evolution of the `attractor' (stable critical point) of the system
for most initial conditions. As we point out in the next section
however the physical significance of this `attractor' should be
interpreted with care since it is an artifact of the allowed
perturbations in the form of the physical law $f(R)$.

The structure of this paper is the following: In the next section we
derive the autonomous system for the cosmological dynamics of $f(R)$
theories. Using as input a particular cosmic history $H(z)$ (eg
$\Lambda$CDM) we show how can this system be transformed so that its
solution provides the dynamics and functional form of $f(R)$. We
also study the dynamics of this transformed system analytically by
deriving its critical points and their stability during the three
eras of the cosmic background history (radiation, matter and
deSitter). We find that there are  `attractor' critical points for
each era which allow an analytical prediction of the dynamics of the
system. We also confirm this analytical prediction by a numerical
solution of the system demonstrating that the evolution of the
system is independent of the initial conditions. In section III we
use the solution of the above system to reconstruct the cosmological
evolution and functional form of the function $f(R)$. We also
demonstrate the agreement between the analytical and numerical
reconstruction of $f(R)$. Finally in section IV we conclude,
summarize and refer to future prospects of this work.

\section{Dynamics of $f(R)$ Cosmologies}

We consider the action (\ref{actfr}) describing the dynamics of
$f(R)$ theories in the Jordan frame \cite{fRpapers}. In the context
of flat Friedman-Robertson-Walker (FRW) universes the metric is
homogeneous and isotropic ie \be \rd s^{2}=-\rd
t^{2}+a^{2}(t)\,\rd{\bf x}^{2}\label{frwmet}\ee and variation of the
action (\ref{actfr}) leads to the following dynamical equations
which are the generalized Friedman equations
\ba 3F H^{2} &=& \rho_{{\rm m}}+\rho_{{\rm rad}}+\frac{1}{2}(FR-f)-3H{\dot  F}\label{fe1} \\
-2F\dot{H} &=& \rho_{{\rm m}}+\frac{4}{3}\rho_{{\rm
rad}}+\ddot{F}-H\dot{F}\label{fe2}\ea where $F\equiv \frac{df}{dR}$
and $\rho_m$, $\rho_{rad}$ represent the matter and radiation energy
densities which are conserved according to \ba
 &  & \dot{\rho}_{{\rm m}}+3H\rho_{{\rm m}}=0\,,\label{mcon}\\
 &  & \dot{\rho}_{{\rm rad}}+4H\rho_{{\rm rad}}=0\,.\label{radcon}\ea
 In order to study the cosmological dynamics implied by equations
 (\ref{fe1}), (\ref{fe2}) we express them as an autonomous system\cite{Amendola:2006we} of
 first order differential equations. To achieve this, we first
 write (\ref{fe1}) in dimensionless form as \be
 1=\frac{\rho_m}{3FH^2}+\frac{\rho_{rad}}{3FH^2} +
 \frac{R}{6H^2}-\frac{f}{6FH^2}-\frac{F'}{F} \label{acon} \ee where
 \be '=\frac{d}{d{\rm
 ln}a}\equiv\frac{d}{dN}=\frac{1}{H}\frac{d}{dt}
 \label{prdef} \ee We now define the dimensionless variables
 $x_1,...,x_4$ as \ba
x_{1} & = & -\frac{F'}{F}\,,\label{x1}\\
x_{2} & = & -\frac{f}{6FH^{2}}\,,\label{x2}\\
x_{3} & = & \frac{R}{6H^{2}}=\frac{H'}{H}+2\,,\label{x3}\\
x_{4} & = & ~\frac{\rho_{{\rm rad}}}{3FH^{2}}=\omr \,.\label{x4}\ea
where in (\ref{x3}) we have used the fact that \be
R=6\left(2H^{2}+\dot{H}\right)=6\left(2H^{2}+H'H\right)\,,\label{R}\ee
and we can associate $x_4$ with $\omr$ and $x_1+x_2+x_3\equiv \ode$
with curvature dark energy (dark gravity). Defining also
$\omms\equiv \frac{\rho_m}{3FH^2}$ we can write equation
(\ref{acon}) as \be \omms=1-x_1-x_2-x_3-x_4 \label{acon1} \ee We may
now use (\ref{prdef}) to express (\ref{fe2}) as \be
\frac{H'}{H}=-\frac{1}{2}\left(\frac{\rho_m}{FH^2}+\frac{4}{3}\frac{\rho_{rad}}{FH^2}+\frac{F''}{F}+\frac{H'}{H}\frac{F'}{F}-\frac{F'}{F}\right)
\label{bau1} \ee or \be x_1'=-1-x_3-3x_2+x_1^2+x_4 \label{au1} \ee
Also, differentiating $x_4$ of (\ref{x4}) with respect to $N$ we
have \be
x_4'=\frac{\rho_{rad}'}{3FH^2}-\frac{\rho_{rad}}{3FH^2}\frac{F'}{F}-\frac{2\rho_{rad}}{3FH^2}\frac{H'}{H}
\label{bau2} \ee or \be x_4'=-2x_3x_4+x_1x_4 \label{au4} \ee where
we have made use of (\ref{radcon}). Similarly, differentiating
(\ref{x2}) with respect to $N$ we find \be
x_2'=\frac{x_1x_3}{m}-x_2(2x_3-x_1-4) \label{au2}\ee where \be
m\equiv \frac{F' R}{f'}=\frac{f,_{R R}R}{f,_R} \label{mdef} \ee and
$,_R$ implies derivative with respect to $R$. Finally
differentiating (\ref{x3}) with respect to $N$ we find \be
x_3'=-\frac{x_1x_3}{m}-2x_3(x_3-2) \label{au3} \ee The autonomous
dynamical system (\ref{au1}), (\ref{au2}), (\ref{au3}), (\ref{au4})
is the general dynamical system that describes the cosmological
dynamics of $f(R)$ theories. It has been extensively studied in Ref.
\cite{Amendola:2006we} for various cases of $f(R)$ (or equivalently
various forms of $m$) and was found to lead to a dynamical evolution
that in most cases is incompatible with observations since it
involves no proper matter era. Some forms of $f(R)$ however were
found to lead to a cosmological evolution that is potentially
consistent with observations. In order to investigate such cases in
more detail we follow a different approach. Instead of investigating
the above autonomous system for various different behaviors of
$m(f(R))$ we {\it eliminate} $m$ from the system by assuming a
particular form for $H(N)$ (ie $x_3(N)$ (see (\ref{x3}))) consistent
with cosmological observations. Once $x_3(N)$ is known we can solve
(\ref{au3}) for $\frac{x_1x_3}{m}$ and substituting in (\ref{au2})
we find \be x_2'=-x_3'-2x_3(x_3-2)-x_2(2x_3-x_1-4) \label{au2a} \ee
which along with (\ref{au1}) and (\ref{au4}) consist a new dynamical
system which is independent of $m$. The study of this system will be
our focus in what follows.

The results of our analysis do not rely on the use of any particular
form of $x_3(N)$ (ie $H(z)$). They only require that the universe
goes through the radiation era (high redshifts), matter era
(intermediate redshifts) and acceleration era (low redshifts). The
corresponding total effective equation of state \be
w_{eff}=-1-\frac{2}{3}\frac{H'(N)}{H(N)} \label{weff}
\ee is \ba w_{eff}&=&\frac{1}{3} \;\;\; {\rm Radiation\; Era} \nn \\ w_{eff}&=&0 \;\;\;\; {\rm Matter\; Era} \label{eraweff} \\
w_{eff}&=&-1 \;\;\; {\rm deSitter\; Era} \nn \ea For the sake of
definiteness however, we will assume a specific form for $H(z)$
corresponding to a $\Lambda$CDM cosmology (\ref{lcdmz}) which in
terms of $N$ takes the form \be H(N)^2=H_0^2 \left[ \omm e^{-3N}+
\orad e^{-4N}+\ol\right] \label{lcdmn} \ee where $N\equiv {\rm ln}
a=-{\rm ln}(1+z)$ and $\ol=1-\omm-\orad$. We can use (\ref{x3}) and
(\ref{lcdmn}) to find $x_3(N)$ as \be x_3(N)=2-\frac{3}{2}\frac{\omm
e^{-3N} +\frac{4}{3}\orad e^{-4N}}{\omm e^{-3N}+ \orad
e^{-4N}+(1-\omm-\orad)} \label{x3lcdm} \ee The crucial generic
properties of $x_3(N)$ are its values at the radiation, matter and
deSitter eras: \ba x_3(N)&=& 0 \;\; \;\; N<N_{rm}
\label{x3rad} \\
x_3(N)&=& \frac{1}{2} \;\; \;\; N_{rm}<N<N_{m\Lambda} \label{x3mat}
\\x_3(N)&=& 2 \;\; \;\; N>N_{m\Lambda} \label{x3ds} \ea

\vspace{0pt}
\begin{table*}
\begin{center}
\caption{The critical points of the system (2.15), (2.21), (2.17)
and their stability in each one of the three eras. Stable points
(attractors) have only negative eigenvalues, saddle points have
mixed sign eigenvalues and unstable points have positive
eigenvalues. \label{table1}}
\begin{tabular}{ccccccc}
\hline \hline\\
\vspace{6pt} \textbf{Era} & \textbf{N Range} \hspace{7pt} & \hspace{7pt} \textbf{$x_1$} \hspace{7pt}& \hspace{7pt} \textbf{$x_2$} \hspace{7pt}& \hspace{7pt} \textbf{$x_3$}  \hspace{7pt}& \hspace{7pt} \textbf{$x_4$} \hspace{7pt}& \hspace{7pt} \textbf{Eigenvalues} \hspace{7pt} \\
\hline
\vspace{6pt} &                  &   -1     &   0    &      0    &    0   &  (3,-2,-1) \\
\vspace{6pt} \textbf{Radiation} & $N<-{\rm ln}\frac{\omm}{\orad} $ & 1  & 0 & 0 & 0 & (5,2,1) \\
\vspace{6pt} $w_{eff}=\frac{1}{3}$& & -4 & 5 & 0 & 0 & (-5,-4,-3)\\
\vspace{6pt} & & 0  & 0 & 0 & 1 & (4,-1,1)\\
\hline
\vspace{6pt} &  &  1  &  -3/8   &  1/2     & -1/8      & (4.386,1,0.114) \\
\vspace{6pt} \textbf{Matter}    & $-{\rm ln}\frac{\omm}{\orad}<N<-\frac{1}{3}{\rm ln}\frac{\ol}{\omm} $  & 0  & -1/2 & 1/2 & 0  & (3.386,-1,-0.886)\\
\vspace{6pt} $w_{eff}=0$& & 0.886 & -0.386 & 1/2 & 0 & (4.272,0.886,-0.114)\\
\vspace{6pt} & & -3.386 & 3.886 & 1/2 & 0 & (-4.386, -4.272, -3.386)\\
\hline
\vspace{6pt} &                 &   0   &  -1   &  2   &   0  & (-4, -3, 1)\\
\vspace{6pt} \textbf{deSitter} & $N>-\frac{1}{3}{\rm ln}\frac{\ol}{\omm} $ & -1 & 0 & 2 & 0  & (-5, -4, -1) \\
\vspace{6pt} $w_{eff}=-1$& & 3  & 0 & 2 & 0  & (4, 3, -1)\\
\vspace{6pt} & & 4  & 0 & 2 & -5 & (5, 4, 1) \\
\hline \hline
\end{tabular}
\end{center}
\end{table*}

\vspace{0pt}
\begin{table*}
\begin{center}
\caption{The `attractor' critical point in each era. \label{table2}}
\begin{tabular}{cccccccccc}
\hline \hline\\
\vspace{6pt} \textbf{Era} & \textbf{N Range} \hspace{7pt} & \hspace{7pt} \textbf{$x_1$} \hspace{7pt}& \hspace{7pt} \textbf{$x_2$} \hspace{7pt}& \hspace{7pt} \textbf{$x_3$}  \hspace{7pt}& \hspace{7pt} \textbf{$x_4$} \hspace{7pt}& \hspace{7pt} \textbf{$w_{eff}$} \hspace{7pt} & \hspace{7pt} \textbf{$\omm$} \hspace{7pt}& \hspace{7pt} \textbf{$\Omega_{DE}$} \hspace{7pt}& \hspace{7pt} \textbf{$\Omega_{rad}$} \hspace{7pt}\\
\hline
\vspace{6pt} \textbf{Radiation} & $N<-{\rm ln}\frac{\omm}{\orad} $ & -4 & 5 & 0 & 0 & 1/3 & 0 & 1 &0 \\
\hline
\vspace{6pt} \textbf{Matter}    & $-{\rm ln}\frac{\omm}{\orad}<N<-{\rm ln}\frac{\ol}{\omm} $ & -3.386  & 3.886  & 1/2 & 0 & 0 & 0 &  1  &0\\
\hline
\vspace{6pt} \textbf{deSitter} & $N>-\frac{1}{3}{\rm ln}\frac{\ol}{\omm} $ & -1 & 0 & 2 & 0  & -1 & 0 & 1  &0\\
\hline \hline
\end{tabular}
\end{center}
\end{table*}

\vspace{0pt}
\begin{table*}
\begin{center}
\caption{The `standard' saddle critical points in each era. These
are also the points producing a linear general relativistic
$f(R)=R-2\Lambda$ (see equation (\ref{grrec1})). \label{table3}}
\begin{tabular}{cccccccccc}
\hline \hline\\
\vspace{6pt} \textbf{Era} & \textbf{N Range} \hspace{7pt} & \hspace{7pt} \textbf{$x_1$} \hspace{7pt}& \hspace{7pt} \textbf{$x_2$} \hspace{7pt}& \hspace{7pt} \textbf{$x_3$}  \hspace{7pt}& \hspace{7pt} \textbf{$x_4$} \hspace{7pt}& \hspace{7pt} \textbf{$w_{eff}$} \hspace{7pt} & \hspace{7pt} \textbf{$\omm$} \hspace{7pt}& \hspace{7pt} \textbf{$\Omega_{DE}$} \hspace{7pt}& \hspace{7pt} \textbf{$\Omega_{rad}$} \hspace{7pt}\\
\hline
\vspace{6pt} \textbf{Radiation} & $N<-{\rm ln}\frac{\omm}{\orad} $ & 0 & 0 & 0 & 1 & 1/3 & 0 & 0  &1\\
\hline
\vspace{6pt} \textbf{Matter}    & $-{\rm ln}\frac{\omm}{\orad}<N<-{\rm ln}\frac{\ol}{\omm} $ & 0  & -1/2 & 1/2 & 0 & 0 & 1 &  0  &0\\
\hline
\vspace{6pt} \textbf{deSitter} & $N>-\frac{1}{3}{\rm ln}\frac{\ol}{\omm} $ & 0 & -1 & 2 & 0  & -1 & 0 & 1  &0\\
\hline \hline
\end{tabular}
\end{center}
\end{table*}

\noindent where $N_{rm}\simeq -{\rm ln}\frac{\omm}{\orad}$ and
$N_{m\Lambda}\simeq -\frac{1}{3}{\rm ln}\frac{\ol}{\omm}$ are the
$N$ values for the radiation-matter and matter-deSitter
transitions. For $\omm=0.3$, $\orad=10^{-4}$ we have $N_{rm}\simeq
-8$, $N_{m\Lambda}\simeq -0.3$. The transition between these eras
is model dependent but rapid and it will not play an important
role in our analysis.

It is straightforward to study the dynamics of the system
(\ref{au1}), (\ref{au2a}), (\ref{au4}) by setting $x_i'=0$ to find
the critical points and their stability in each one of the three
eras corresponding to (\ref{x3rad})-(\ref{x3ds}). Notice that even
though this dynamical system is not autonomous at all times it can
be approximated as such during the radiation, matter and deSitter
eras when $x_3$ is approximately constant. The critical points and
their stability are shown in Table I. 

The stability analysis of Table I assumes that $x_3=const$ and
therefore it is not identical to the full stability analysis where
$x_3$ would be allowed to vary. The usual stability analysis of
cosmological dynamical systems assumes a particular cosmological
model (eg a form of $f(R)$ or $m$) and in the context of this
`physical law', the stability of cosmic histories $H(N)$ is
investigated. In this context clearly a stable cosmic history is
the one preferred by the model.

In the reconstruction approach however the stability analysis has a
very different meaning. Here we do not fix the model $f(R)$
(`physical law'). Here we fix the cosmic history and allow the
physical law $f(R)$ to vary in order to predict the required cosmic
history. Thus our stability analysis concerns the `physical law'
$f(R)$ and not the particular cosmic history. Since the physical law
is usually fixed by Nature the instabilities we find are not
physically relevant but they are only useful to understand
analytically the phase space trajectories we obtain numerically. The
physically interesting quantities are the values of the critical
points we find in each era in the context of the $\Lambda CDM$
cosmic history. These tell us the possible physical laws $f(R)$ that
can reproduce a $\Lambda CDM$ cosmic history. As shown in Table I,
one of these laws is clearly the general relativistic
$f(R)=R-2\Lambda$.

The important point to observe in Table I is that in each era there
are four critical points one of which correspond to the general
relativistic $f(R)=R-2\Lambda$. Three of the four critical points in
each era are not stable. This however does not imply that these
points are not cosmologically relevant. These instabilities are not
instabilities of the trajectory $H(N)$ (which we keep fixed) but of
the forms of $f(R)$ which is allowed to vary. Thus they are not so
relevant physically since in a physical context $f(R)$ is assumed to
be fixed a priori. The `attractor' critical points of Table I are
relevant only for technical reasons since they allow a comparison
between a numerical evolution and an analytical prediction of the
evolution of the system. In a more realistic situation where the
perturbations of the `physical law' $f(R)$ would be turned off, all
critical points would correspond to valid reconstructions of a
$\Lambda CDM$ cosmic history.

If we allow for $f(R)$ perturbations (but not of $H(N)$
perturbations), the evolution of the system is determined by just
following the evolution of the `attractors' of Table I through the
three eras. This evolution is presented in Table II showing the
`attractors' in each era. We stress however that this is not
necessarily a preferred cosmological trajectory for the reasons
described above. 

\begin{figure}[t!]
\hspace{0pt}\rotatebox{0}{\resizebox{.5\textwidth}{!}{\includegraphics{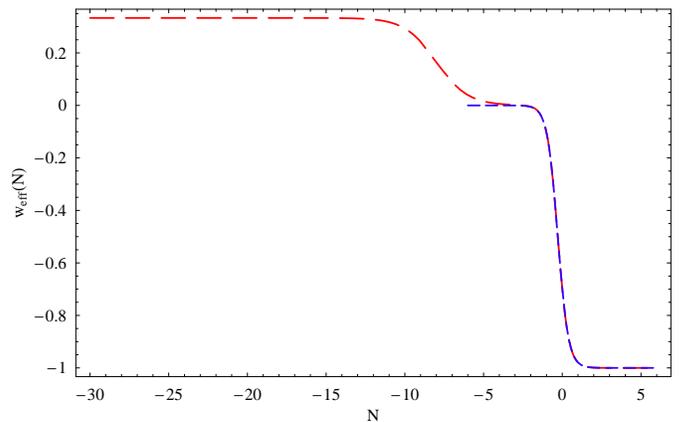}}}
\vspace{-20pt}{\caption{The effective equation of state
$w_{eff}(N)$ imposed on the dynamical system (obtained from
(\ref{weff}) using (\ref{lcdmn})). The long dashed red line starts
deep in the radiation era while the short dashed blue line starts
in the matter era and ignores radiation ($\orad=0$).}}
\label{fig1}
\end{figure}

The `standard' critical points are shown in Table III and they are
{\it the only} critical points that have in addition to the
correct expansion rate properties, the required values of
$(\omr,\omms,\ode)$ in each era. As shown in the next section
these saddle critical points reconstruct the general relativistic
$f(R)$ ie $f(R)=R-2\Lambda$. It is therefore clear that {\it
nonlinear} $f(R)$ theories can produce an observationally
acceptable cosmic history but not with the required values of
$(\omr,\omms,\ode)$ in each era. We should stress that our
analysis has not excluded the possibility of physical values of
$(\orad,\omm,\ode)$ in the case of cosmic histories oscillating
around the anticipated $w_{eff}$ in each era or a $w_{eff}$ that
is continuously evolving. These special cases however maybe
severely constrained observationally.

To confirm the dynamical evolution implied by the `attractors' of
Table I, we have performed a numerical analysis of the dynamical
system (\ref{au1}), (\ref{au2a}), (\ref{au4}) using the ansatz
(\ref{x3lcdm}) for $x_3$ with $\omm=0.3$ and $\orad=10^{-4}$. This
ansatz for $x_3(N)$ leads to the $w_{eff}(N)$ shown in Fig. 1. We
have set up the system initially, close to the `standard'
radiation era saddle point $(0,0,0,1)$ and allowed it to evolve.
As seen in Fig. 2 and Fig. 3  much before the onset of the matter
era ($N\equiv N_{rr}\simeq -25<-8\simeq N_{rm}$) the slow (but
non-zero) evolution of $x_3(N)$ forces the phase space trajectory
to depart from the saddle point $(0,0,0,1)$ and head towards the
radiation era stable `attractor' $(-4,5,0,0)$ where it stays
throughout the rest of the radiation era ($w_{eff}\simeq
\frac{1}{3}$).


\begin{figure*}[ht!]
\rotatebox{0}{\resizebox{1\textwidth}{!}{\includegraphics{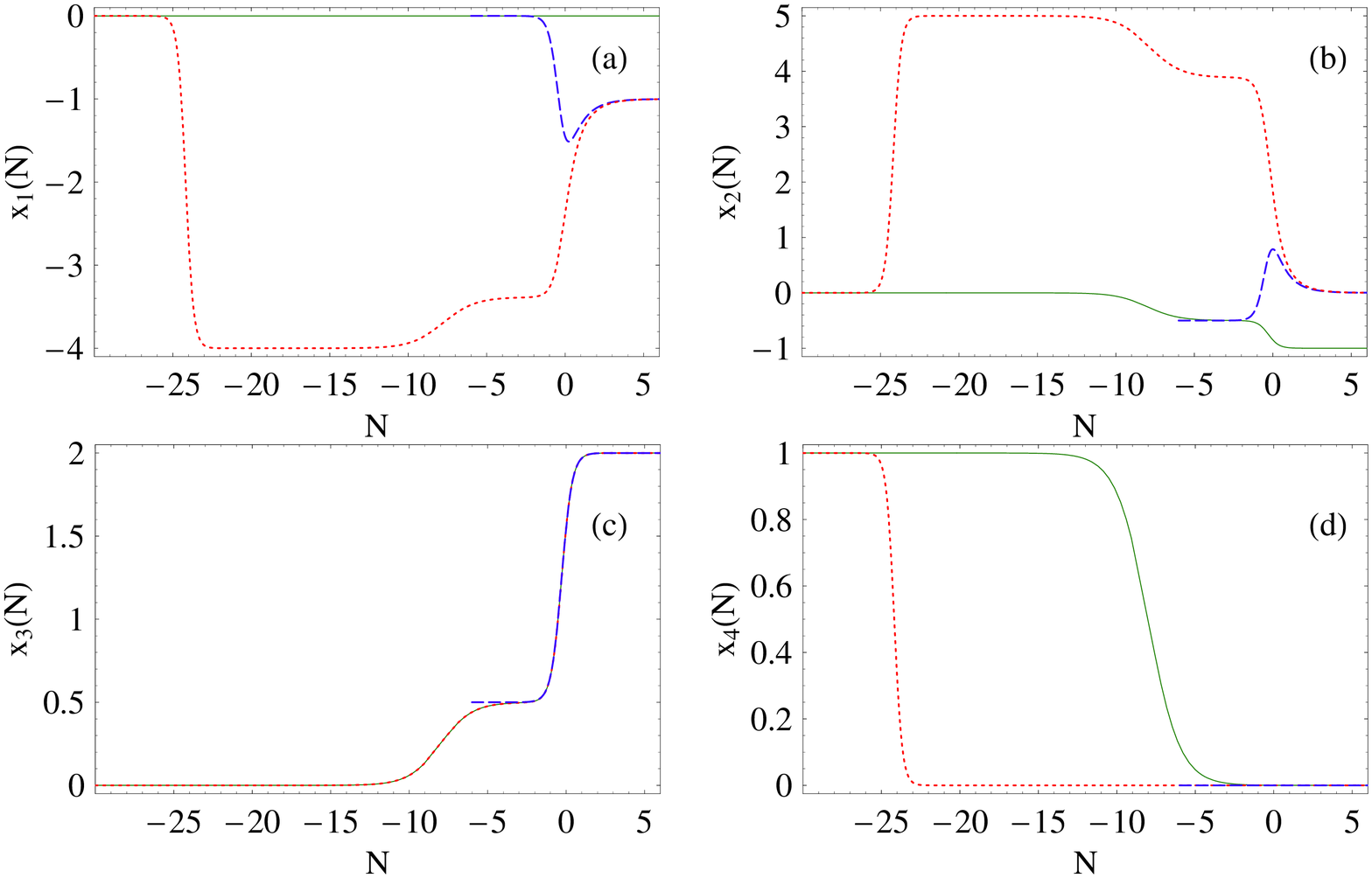}}}
\vspace{0pt}{ \caption{The evolution of the variables $x_1(N)$,
$x_2(N)$, $x_3(N)$ and $x_4(N)$ for `standard' radiation era initial
conditions (dotted red line) and `standard' matter era initial
conditions (dashed blue line). The perturbed trajectories are
rapidly dragged by the stable `attractors' of each era. The
numerically obtained evolution along the `standard' saddle points of
Table III is also shown (continuous green line). The instabilities
of this trajectory are bypassed by using the constrained system
(\ref{lcdm1})-(\ref{lcdm2}) instead of the full system (\ref{au1}),
(\ref{au2a}), (\ref{au3}).}} \label{fig2}
\end{figure*}

\begin{figure*}[hb!]
\rotatebox{0}{\resizebox{1\textwidth}{!}{\includegraphics{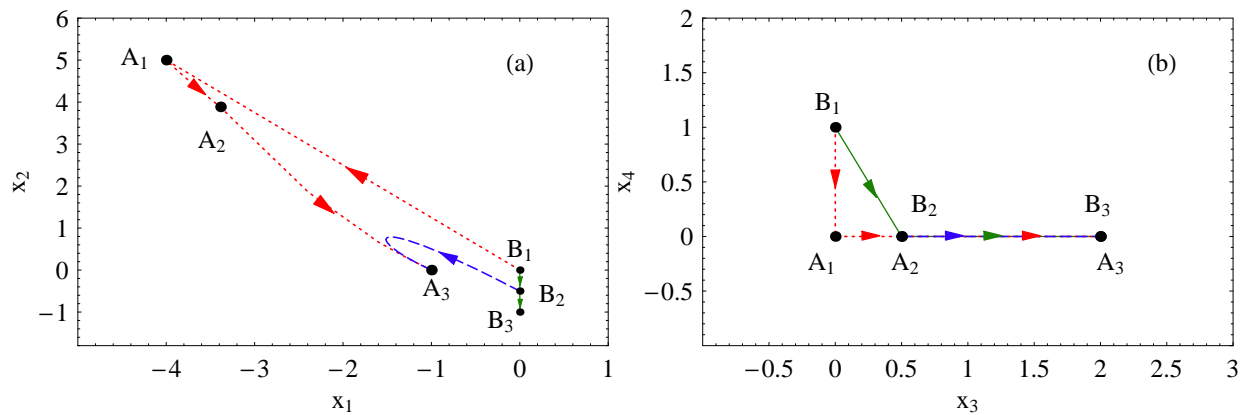}}}
\vspace{0pt}{ \caption{The phase space trajectories on  the $x_1 -
x_2$  plane (Fig3a) and $x_3 - x_4$ plane (Fig3b) for `standard'
radiation era initial conditions (dotted red line) and `standard'
matter era initial conditions (dashed blue line). The trajectory
corresponding to the numerically obtained evolution along the saddle
points of Table III is also shown (continuous green line). The
points $A_1$, $A_2$, $A_3$ correspond to the `attractors' of each
era (radiation, matter and deSitter respectively) while the points
$B_1$, $B_2$, $B_3$ correspond to the `standard' critical points of
each era (see Tables II and III). Notice that on the projection of
Fig. 3b  the `attractor' points $A_2$, $A_3$ coincide with the
`standard' critical points $B_2$, $B_3$.}} \label{fig3}
\end{figure*}
\clearpage

Subsequently, when $x_3(N)$ enters the matter era ($w_{eff}=0$) at
$N_{rm}\simeq -8$, the trajectory follows the evolution of the
`attractor' fixed point and heads towards the matter era
`attractor' $(-3.386,3.886,0.5,0)$ ignoring the saddle point
$(0,-1/2,1/2,0)$ of the `standard' matter era. Finally when the
matter era is over, the trajectory heads towards the deSitter
`attractor' $(-1,0,2,0)$ which is also distinct from the
`standard' deSitter saddle point $(0,-1,2,0)$. Notice that the
deSitter `attractor' is inconsistent with observations due to the
implied large variation of the effective Newton's constant
$G(N)=\frac{1}{F(N)}$ even though this inconsistency could be
ameliorated by `chameleon' type mechanisms \cite{Khoury:2003rn}.
The evolution of $(\omr,\omms,\ode)$ corresponding to the phase
space trajectories of Figs. 2 and 3 is shown in Fig. 4a. Notice
that $\omms=0$ throughout the `attractor' evolution of the system
and the $w_{eff}=0$ of the matter era is induced by curvature dark
gravity excitations.

We have also tested initial conditions in the matter era starting
the evolution on the saddle point $(0,-\frac{1}{2},\frac{1}{2},0)$
corresponding to the `standard' matter era. In this case we also
ignore radiation setting $\orad=0$. We get an evolution of the
system (see Figs 2, 3, 4b) which stays on the `standard' matter
era $\omms=1$ for about 3 expansion times but before the onset of
the acceleration era it gets absorbed by the `attractor' towards
the nonstandard deSitter critical point $(-1,0,2,0)$.

The above evolution along the `attractor' critical points is a
result of the `physical law' $f(R)$ perturbations.  We can also
reproduce trajectories that go through critical points that are
not stable by turning off these perturbations. For example we can
recover the saddle critical point sequence \be
(0,0,0,1)\rightarrow (0,\frac{1}{2},-\frac{1}{2},0)\rightarrow
(0,-1,2,0) \label{grseq} \ee by fixing $x_1=0$ in the system
(\ref{au1}), (\ref{au2a}), (\ref{au4}) and reducing it to the
system \ba -1-x_3 -3x_2 +x_4 &=& 0 \label{lcdm1} \\ -2 x_3 x_4&=&
x_4' \label{lcdm2} \ea which can be easily solved using the
ansatz (\ref{x3lcdm})

\begin{figure*}
\begin{center}
\rotatebox{0}{\resizebox{1\textwidth}{!}{\includegraphics{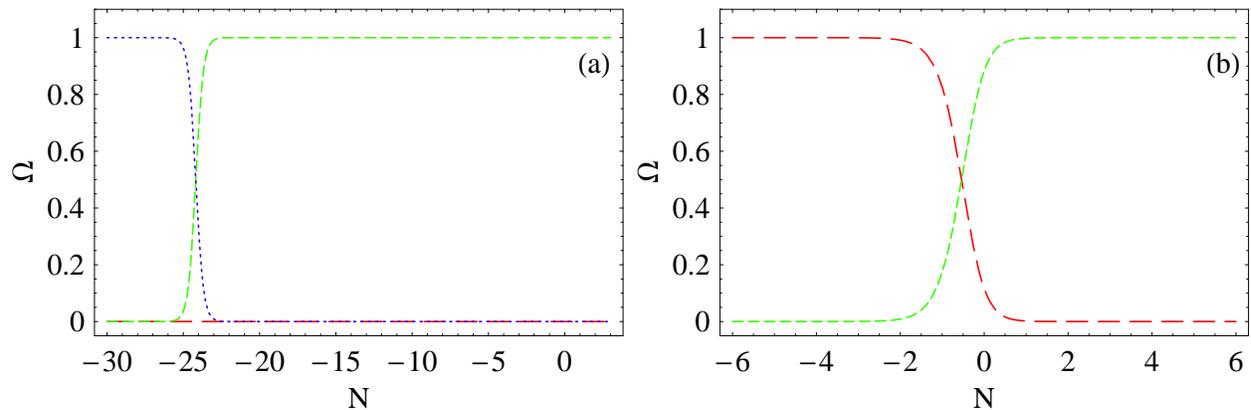}}}
\vspace{0pt}{ \caption{The evolution of $\Omega_{m}$ (long-dashed),
$\Omega_{rad}$ (dotted) and $\Omega_{DE}$ (short-dashed) parameters
for `standard' radiation era initial conditions (Fig4a) and
`standard' matter era initial conditions (Fig4b). In both cases the
`attractor' critical points of Table II rapidly take over and drag
the trajectories.}} \label{fig4}
\end{center}
\end{figure*}

\noindent(see Fig. 2 and Fig. 3). The alternative approach of
solving the decoupled pair (\ref{au2}), (\ref{au4}) does not lead to
the correct result because the $f(R)$ perturbations are not turned
off and the constraint is not respected in this case.

It is straightforward to reconstruct the functions $f(R)$ that
correspond to the saddle general relativistic trajectory
(\ref{grseq}) and to the `attractor' sequence of Table II. The
functional forms of $f(R)$ may also be reconstructed on any one of
the critical points of Table I. These tasks are undertaken in the
next section.

\section{Reconstruction of $f(R)$}

We now reconstruct the form of the function $f(R)$ that corresponds
to each one of the critical points of the system shown in Table I.
This reconstruction is effectively an approximation of $f(R)$ in the
neighborhood of each critical point. It is particularly useful
because most of the dynamical evolution takes place close to the
fixed points. Consider a critical point of the form $({\bar
x}_1,{\bar x}_2,{\bar x}_3,{\bar x}_4)$. Using (\ref{x1}) we find
\be F=F_0 e^{-{\bar x}_1 N} \label{fsol1} \ee where $F_0$ is a
constant. We may eliminate $N$ in favor of $R$ using the input form
of $H(N)$ (equation (\ref{lcdmn}) in (\ref{R}) to obtain (setting
$H_0^2=1$) \be R(N)=3 \left[4\ol + \omm e^{-3N}\right] \label{rn}
\ee which leads to \be F=F_0
\left(\frac{R-12\ol}{3\omm}\right)^{\frac{{\bar x}_1}{3}}
\label{fsol2} \ee and by integration we get \be
f(R)=\frac{3F_0(3\omm)^{-\frac{{\bar x}_1}{3}}(R-12\ol)^{\frac{{\bar
x}_1}{3}+1}}{{\bar x}_1+3}+f_0 \label{ffsol1} \ee where $f_0$ is an
integration constant. Expressing (\ref{ffsol1}) in terms of $N$
using (\ref{rn}) we obtain \be f(N)=\frac{9F_0 \omm e^{-({\bar x}_1
+3)N}}{{\bar x}_1+3} +f_0 \label{ffsol2} \ee It is now
straightforward to use the expressions for $f(N)$  $R(N)$ and $H(N)$
to find $x_2(N)$ (equation (\ref{x2})), $x_3(N)$ (equation
(\ref{x3lcdm})) and $x_4(N)$ (equation (\ref{x4})).  We thus find
\be x_2(N)=-\frac{\left(\frac{3\omm}{2({\bar
x}_1+3)}e^{-3N}+\frac{f_0}{6 F_0}e^{{\bar x}_1 N}\right)}{\omm
e^{-3N} + \orad e^{-4N} +\ol} \label{x2ncrit} \ee and \be
x_4(N)=\frac{\orad e^{({\bar x}_1-4)N}}{F_0 (\omm e^{-3N} + \orad
e^{-4N} +\ol)} \label{x4ncrit} \ee while $x_3(N)$ is given by
(\ref{x3lcdm}). Using equations (\ref{x2ncrit}), (\ref{x3lcdm}) and
(\ref{x4ncrit}) we may verify the ${\bar x}_2$, ${\bar x}_3$, ${\bar
x}_4$ values of each critical point by considering the appropriate
range of $N$ in each era and the corresponding value of ${\bar
x}_1$. By demanding consistency with the values of Table I we may
obtain the values of the constants $f_0$ and $F_0$.

As an example let's consider the sequence (\ref{grseq})
corresponding to the `standard' cosmological eras (Table III). It is
easy to see, using ${\bar x}_1=0$ and the appropriate range of $N$
in (\ref{x2ncrit}), (\ref{x3lcdm}) and (\ref{x4ncrit}) that we
obtain the correct values for ${\bar x}_2$, ${\bar x}_3$, ${\bar
x}_4$ in the radiation and matter eras for any value of $F_0$,
$f_0$. In the deSitter era ($N>>1$) the value of $f_0$ needs to be
fixed to get agreement with ${\bar x}_2=-1$ of Table I. In
particular from (\ref{x2ncrit}) we find \be {\bar
x}_2=-\frac{f_0}{6F_0 \ol}=-1 \label{f0det1} \ee which implies \be
f_0=6 F_0 \ol \label{f0det2} \ee Using now (\ref{f0det2}) and
setting ${\bar x}_1=0$ in (\ref{ffsol1}) we reconstruct the expected
result \be f(R)=F_0(R-6\ol) \label{grrec1} \ee which is valid for
all three eras since in this sequence the value of ${\bar x}_1$
remains constant. In a similar way we may reconstruct $f(R)$ for any
critical point in one of the three eras.

\begin{figure*}
\rotatebox{0}{\resizebox{1\textwidth}{!}{\includegraphics{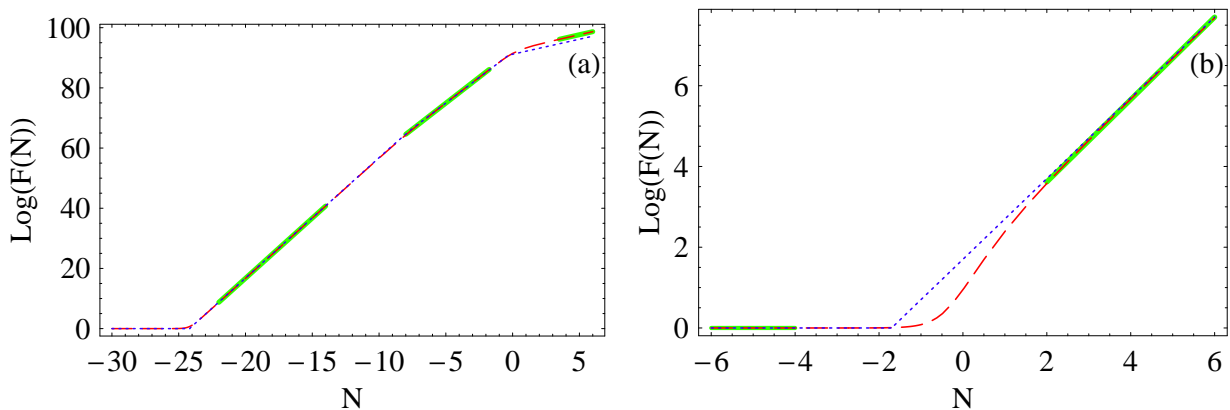}}}
\vspace{0pt}{ \caption{The form of $log(F(N))$ in the numerical
reconstruction (dashed lines) and its analytical approximation
using both the interpolating steps 1-3 (dotted lines) and the
application of the analytical expression (\ref{ffsol2}) valid in
each era (thick green lines). The agreement between the three
approaches is very good. 5a: `standard' radiation era initial
condition, 5b: `standard' matter era initial condition.}}
\label{fig5}
\end{figure*}

\begin{figure*}
\rotatebox{0}{\resizebox{1\textwidth}{!}{\includegraphics{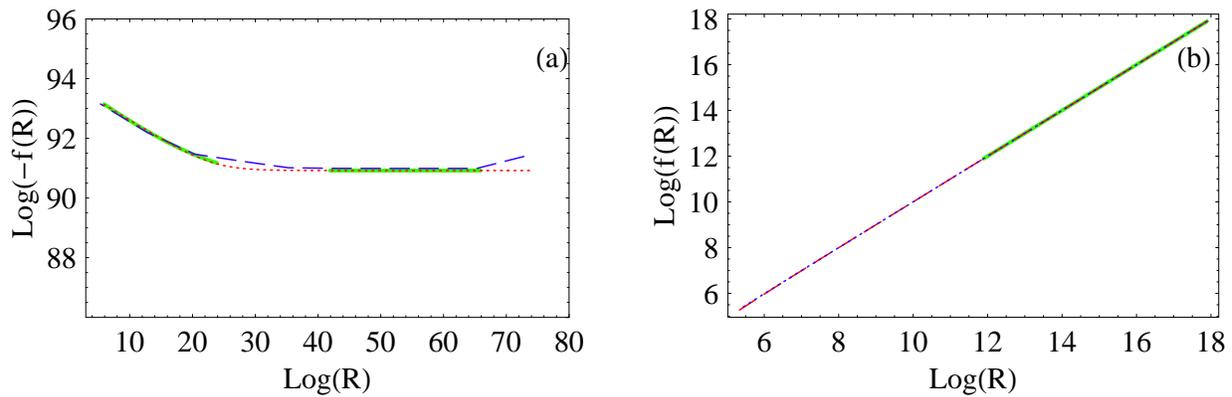}}}
\vspace{0pt}{ \caption{The form of $f(R)$ in the numerical
reconstruction (dashed lines) and its analytical approximation
using both the interpolating steps 1-3 (dotted lines) and the
application of the analytical expression (\ref{ffsol1}) valid in
each era (continuous green lines). 6a: `standard' radiation era
initial condition with the continuous green lines corresponding to
radiation era (larger $R$) and matter era (smaller $R$). The
deSitter era is not shown since it corresponds to a single point
(constant $R$). 6b: `standard' matter era initial condition with
the continuous green line corresponding to matter era. The
deSitter era is not shown since it corresponds to a point
(constant $R$).}} \label{fig6}
\end{figure*}
We have therefore extended previous studies showing that $f(R)$
theories can only be viable in very restricted cases by showing
that even these restricted cases can not reproduce a viable
$\Lambda$CDM cosmology where $w_{eff}$ is constant during the
matter and radiation eras and $\orad$, $\omms$ take their
cosmologically anticipated values. It therefore becomes clear that
if the accelerating expansion of the universe is due to physics in
the gravitational sector it may probably have to be a more general
theory than $f(R)$ modified gravity. Such a theory could very well
be scalar-tensor gravity (or equivalently coupled dark
energy\cite{coupled}) whose cosmological dynamical properties and
constraints need to investigated in detail.

In the case of sequence of transitions among critical points which
involve different values of ${\bar x}_1$ the reconstruction can be
done by either numerical determination of $x_1(N)$ or by
approximating it as a sequence of step functions. For example, the
steps involved in the reconstruction of the `attractor' trajectory
shown in Fig. 2 and Fig. 3 are the following:

\begin{enumerate}
\item Use (\ref{x1})
along with the numerical solution $x_1(N)$ to find the function
$F(N)=f,_R (N)$ as \be F(N)=F_0 e^{-\int_{N_{min}}^N x_1(N')dN'}
\label{fnanal} \ee  The numerical solution $x_1(N)$ of Fig. 2a can
be approximated as a piecewise constant function with values
determined by the corresponding `attractors' of each cosmological
era and by the initial conditions ie \ba x_1(N)&=&0 \;\;\; -30<N<N_{rr} \label{x1na} \\
x_1(N)&=&-4 \;\;\; N_{rr}<N<N_{rm} \label{x1nb} \\
x_1(N)&=&-3.386 \;\;\; N_{rm}<N<N_{m\Lambda} \label{x1nc} \\
x_1(N)&\simeq&-1 \;\;\;\; N_{m\Lambda}<N \label{x1nd} \ea (where
$N_{rr}\simeq -25$) thus leading to an analytical approximation for
$F(N)$. The resulting form of $\ln (F(N))$ in both the numerical
reconstruction and its analytical approximation is shown in Fig. 5
(5a: `standard' radiation era initial condition, 5b: `standard'
matter era initial condition).
\item Use equation (\ref{x2}) to find
$f(N)$ from $F(N)$ ie \be f(N)=-6 x_2(N) F(N) H(N)^2 \label{fnanal1}
\ee where $H(N)$ is given by (\ref{lcdmn}), $x_2(N)$ is numerically
obtained and shown in Fig. 2b and $F(N)$ is obtained in the previous
step. As in the case of $x_1(N)$, $x_2(N)$ can be analytically
approximated as \ba x_2(N)&=&0 \;\;\; -30<N<N_{rr} \label{x2na} \\
x_2(N)&=&5 \;\;\; N_{rr}<N<N_{rm} \label{x2nb} \\
x_2(N)&=&3.886 \;\;\; N_{rm}<N<N_{m\Lambda} \label{x2nc} \\
x_2(N)&\simeq&0 \;\;\;\; N_{m\Lambda}<N \label{x2nd} \ea using the
corresponding `attractors' to obtain an analytical expression for
$f(N)$. \item The resulting form of $f(N)$ can then be combined with
equation (\ref{rn}) for $R(N)$ to reconstruct the function $f(R)$.
The resulting form of $f(R)$ is shown in Fig. 6 for both the
numerical reconstruction and its analytical approximation (6a:
radiation era initial conditions, 6b: matter era initial
conditions). \end{enumerate} We can fit the reconstructed $f(R)$ of
Figs. 6a and 6b to the analytic form of equation (\ref{ffsol1}) for
each era respectively so as to find the parameters $F_0$, ${\bar
x}_1$ and $f_0$. The results are shown in Table IV.

\begin{table}
  \centering
  \caption{ The parameters $F_0$,
${\bar x}_1$ and $f_0$ for the reconstructed $f(R)$ of Figs. 6a and
6b .}\label{table4}
  \begin{tabular}{ccccc}
\hline \hline\\
   \textbf{Radiation Era cond.}  & ${\bar x}_1$ & $Log(F_0)$ & $Log(-f_0)$ &  \\
    Radiation Era         & -3.99 & 96.73 & 90.92 &  \\
    Matter Era            & -3.45 & 92.17 & 90.59 &  \\
    deSitter era          & -1.01 & 92.62 & 94.03 &  \\
    \\
 \hline\\

    \textbf{Matter Era cond.} & ${\bar x}_1$ & $F_0$ & $f_0$ &  \\
    Matter Era               & 0    & 1    & -1.51   &  \\
    deSitter era            & -1.02 & 4.83 & -16.56 &  \\
 \hline \hline\\
  \end{tabular}
\end{table}

Notice that the best fit values of ${\bar x}_1$ coincide with the
corresponding `attractor' critical points of Table II as expected.
This verifies the validity of the reconstructed $f(R)$ expression
from (\ref{ffsol1}). A similar reconstruction analysis can be
performed for any other sequence of critical points. As discussed in
section II any such sequence is equally interesting cosmologically
since the existence of the `attractor' is an artifact of the $f(R)$
perturbations.

\section{Conclusion-Outlook}
We have shown analytically and numerically that nonlinear $f(R)$
gravity theories can reproduce the background expansion history
$H(z)$ indicated by observations even when $f(R)$ does not reduce to
general relativity at early times. In that case the universe gets
dominated by dark gravity during its evolution as opposed to
radiation or matter. This result relies on the values of all the
critical points we found assuming only that the radiation era
corresponds to a {\it constant} effective equation of state
parameter $w_{eff}=\frac{1}{3}$ while for the matter era we have
$w_{eff}=0$.

Our analysis indicates $f(R)$ models can be viable if $f(R)$
deviates from general relativity at early times. Thus a viable
$f(R)$ theory must satisfy one of the following:
\begin{itemize}
\item Either $f(R)$ reduces to general relativity at early times, but
departs from general relativity at late times (a well known
case\cite{fRpapers}).\item Or dark gravity in the forms derived in
our paper mimics radiation or matter at both the background level
and the perturbative level. The later would clearly require a
separate analysis of perturbations of the model.
\end{itemize}

{\bf Numerical Analysis:} The mathematica files with the
numerical analysis of this study may be found at
http://leandros.physics.uoi.gr/frlcdm/frlcdm.htm or may be sent by
e-mail upon request.

{\bf Acknowledgements:} This work was supported by the European
Research and Training Network MRTPN-CT-2006 035863-1 (UniverseNet).
S.F. is supported by a Marie Curie Intra-European Fellowship of the
European Union (contract number MEIF-CT-2005-515028). S.N.
acknowledges support from the Greek State Scholarships Foundation
(I.K.Y.).


\begin{thebibliography}{99}
\bibitem{ftn} This type of evolution would be more
likely if we had found complex eigenvalues for the saddle point
$(0,-\frac{1}{2},\frac{1}{2},0)$ which would allow for oscillations
around the correct matter era.
\end{thebibliography}
\end{document}